\documentclass[pra,twocolumn,showpacs,preprintnumbers,amsmath,amssymb]{revtex4-1}
\usepackage{amssymb,amsmath}
\usepackage{color}
\usepackage{graphicx}
\usepackage{dcolumn}
\usepackage{bm}
\usepackage{latexsym,epsfig}

\begin{document}
\preprint{\today}

\title{Phases, transitions, and patterns in the one-dimensional Extended Bose-Hubbard model}

\author{Jamshid Moradi Kurdestany}
\email{jmkurdestany@gmail.com} \affiliation{Centre for
Condensed Matter Theory, Department of Physics, Indian Institute of
Science, Bangalore 560 012, India}\altaffiliation[Also
at~]{Jawaharlal Nehru Centre For Advanced Scientific Research,
Jakkur, Bangalore, India}
\author{Ramesh V. Pai}
\email{rvpai@unigoa.ac.in} \affiliation{Department of Physics, Goa
University, Taleigao Plateau, Goa 403 206, India}
\author{Subroto Mukerjee}
\email{smukerjee@physics.iisc.ernet.in
} \affiliation{Centre for
Condensed Matter Theory, Department of Physics, Indian Institute of
Science, Bangalore 560 012, India}\altaffiliation[Also
at~]{Centre for Quantum Information and Quantum Computing, Indian Institute of Science,
Bangalore 560 012, India}

\author{Rahul Pandit}
\email{rahul@physics.iisc.ernet.in} \altaffiliation[Also
at~]{Jawaharlal Nehru Centre For Advanced Scientific Research,
Jakkur, Bangalore, India} \affiliation{Centre for Condensed Matter
Theory, Department of Physics, Indian Institute of Science,
Bangalore 560012, India.}
\date{\today}
\begin{abstract}
We carry out an extensive study of the phase diagram of the
extended Bose Hubbard model, with a mean filling of one boson per
site, in one dimension by using the density matrix
renormalization group and show that it can have Superfluid (SF), Mott-insulator (MI), density-wave (DW) and Haldane-insulator (HI) phases depending on the precise value of filling and how edge states are handled. We show that the critical exponents and central charge for the HI-DW, MI-HI
and SF-MI transitions are consistent with those for models in the
two-dimensional Ising, Gaussian, and Berezinskii-Kosterlitz-Thouless (BKT) universality classes, respectively; and we suggest that the SF-HI transition may be more exotic than a simple BKT transition.
\end{abstract}

\maketitle
\section{Introduction}

Over the last two decades, experiments on cold alkali atoms in traps have
obtained various phases of correlated bosons and fermions~\cite{reviews44}.
Microscopic interaction parameters can be tuned in such experiments, which
provide, therefore, excellent laboratories for studies of quantum phase
transitions, such as those between superfluid (SF) and Mott-insulator (MI)
phases in a system of interacting bosons in an optical
lattice~\cite{fisher8944,sheshadri44,jaksch44}. The SF and MI phases are,
respectively, the protypical examples of gapless and gapped phases in such
systems; in addition, it may be possible to obtain other phases, e.g.,
density-wave (DW)~\cite{kuhner9844, rvpai0544, entspect44, kurdestany1244},
Haldane-insulator (HI)~\cite{torre0644,berg0844}, and supersolid
(SS)~\cite{kovrizhin0544} phases, in systems with a dipolar condensate of
$^{52}{\rm Cr}$ atoms~\cite{werner0544}. The first step in developing an
understanding of such experiments is to study lattice models of bosons with
long-range interactions~\cite{goral0244}; the simplest model that goes beyond the Bose-Hubbard model with onsite, repulsive interactions~\cite{fisher8944,sheshadri44} is the extended Bose-Hubbard model
(EBHM), which allows for repulsive interactions between bosons on a site and
also on nearest-neighbor sites~\cite{rvpai0544,kurdestany1244}. The
one-dimensional (1D) EBHM is particularly interesting because (a) quantum
fluctuations are strong enough to replace long-range SF order by a
quasi-long-range SF, with a power-law decay of order-parameter
correlations~\cite{rvpai0544}, and (b) even a slight deviation from integer filling and the state of the boundaries play important roles here, in so far as they can modify the phase
diagram~\cite{torre0644,berg0844,deng1244,rossini1244}, in a way that is, at first glance, not possible in the thermodynamic limit~\cite{thermodynamiclimit44}.

The 1D EBHM is defined by the Hamiltonian
\begin{equation} {\cal
H} = -t\sum_i (a_{i}^{\dagger} a_{i+1} + H.c)+\frac{U}{2}\sum_i
{\hat n}_i ({\hat n}_i -1)+V\sum_i {\hat n}_i {\hat n}_{i+1},
\label{eq:ebhmodel}
\end{equation}
where $t$ is the amplitude for a boson to hop from site $i$ to
its nearest-neighbor sites, $H.c.$ denotes the Hermitian
conjugate, $a_i^{\dag}, \, a_i$, and ${\hat n}_i \equiv
a_i^{\dag} a_i$ are, respectively, boson creation, annihilation,
and number operators at the site $i$, the repulsive potential
between bosons on the same site is $U$, and $V$ is the repulsive
interaction between bosons on nearest-neighbor sites.

At a filling $\rho$ of one boson per site, the phases of this
model can be understood simply in the limits $(U/t, V/t)
\rightarrow 0$ and $(U/t, V/t) \rightarrow \infty$.  In the
former limit, the hopping energy scale dominates over the
interaction scales; this causes the bosons to delocalize and an
SF phase is obtained.  In the opposite limit, the interaction
scales dominate. Two possibilities exist in this strong-coupling
limit: (a) If $U > V$, the system prefers to have one boson at
every site to minimize the energy costs of double and multiple
occupancy, so an MI phase is stabilized; this has a gap (a charge
gap if the bosons are charged) and a uniform average filling of
one boson per site. (b) If $V > U$, the system prefers to have a
non-uniform filling, with an average of two and zero bosons on
adjacent sites, to minimize the nearest-neighbor-interaction
energy, i.e., the system has a density-wave (DW) phase with wave
vector $k=\pi$. These phases have been obtained in an early
density-matrix-renormalization-group (DMRG)~\cite{rvpai0544}
study of the EBHM.


We now give a qualitative argument, first introduced in
Ref.~\cite{torre0644} for an EBH model with long-range
interactions, which suggests that, at intermediate values of $U$
and $V$, an exotic phase with a string order parameter can also
be obtained. In the strong-coupling regime, the MI and DW states
obtained are both, approximately, eigenstates of the
site-occupation number operators for bosons; thus, they have very
small boson-number fluctuations. In the intermediate-coupling
range, we expect the magnitude of these fluctuations to increase.
At an average filling of one boson per site, the lowest-energy
number fluctuations, over the MI state, have zero or two bosons
per site; similarly, the lowest-energy number fluctuations, over
the DW state, have one boson per site. Thus, the
intermediate-coupling-regime state, which is obtained regardless
of whether we start from the MI or DW phase, is one with three
possible occupancies per site, namely, $0$, $1$, and $2$ bosons
per site. We can, therefore, use an effective, spin-one model for
this system at intermediate values of $U$ and $V$, which are not
so large that a spin-$1/2$ description is adequate and not so
small that there is a significant probability of more than $2$ bosons at a site.

In the paper in which Dalla Torre, {\it et al.}~\cite{torre0644}
have introduced this line of reasoning, in the context of a model
of bosons with long-range interactions, they have shown,
furthermore, that the effective, spin-1 model admits a
Haldane-type phase, namely, the Haldane Insulator (HI). A Haldane
phase, which arises in chains of integer-valued
spins~\cite{FDMHaldane}, is characterized by a gap and a nonzero
string order parameter~\cite{AROVAS}. The original work of
Haldane~\cite{FDMHaldane} has used spin systems with $SU(2)$
symmetry; it has been shown subsequently that the Haldane phase,
in this case, is a state with a broken, hidden $Z_2 \times Z_2$
symmetry, and this phase can exist even in the absence of $SU(2)$
symmetry in the Hamiltonian~\cite{KENNEDY}. The robustness of the
Haldane phase to various types of perturbations has been
investigated recently in Ref.~\cite{FPollmann}.

When we map the EBH model, in the intermediate-coupling range,
onto an effective, spin-1,one model, the latter does not have
$SU(2)$ symmetry because of the extended-interaction term $V$.
Nevertheless, we expect the Haldane phase to survive;
and we demonstrate its existence and elucidate the conditions
under which it is obtained.

The intermediate-coupling regime can also be studied by
proceeding from the SF phase, which has very strong number
fluctuations; these fluctuations are suppressed as the strength
of the interactions is increased. In this case it might be
possible to stabilize an SS phase, in which number fluctuations
are not sufficiently large to localize the bosons completely (and
thus destroy superfluidity), but strong enough to produce a
modulation in average density.  The existence of the SS phase in
the EBH model needs a very careful study as we explain in detail
below. However, there have been claims of the existence of an SS
phase in the EBH model~\cite{entspect44}.

In addition to identifying the various phases of the EBH model,
we must also characterize the transitions between the different
phases. The broken symmetries of these phases can, in principle,
be used to guess the possible universality classes of the
continuous transitions (as in the Landau
paradigm~\cite{paradigm}). However, it is extremely important to
obtain these universality classes by direct studies, which must
often be numerical, for the following two reasons: (1) Strong
quantum fluctuations in one dimension can modify the natures of
the transitions that might be expected on the basis of the
Landau-paradigm view; and (2) energy considerations can make a
first-order transition preempt a continuous transition, which we
might expect na\"ively in a Landau picture.

The broken symmetries in the SF and DW phases are, respectively,
$U(1)$ (guage) and $Z_2$ (lattice translation). The MI state
breaks no symmetries; and the Haldane (HI) phase has a hidden,
broken symmetry and no local order parameter. Thus, we expect the
following universality classes: (A) The SF-MI transition to be in
the $U(1)$ universality class appropriate for a
one-dimensional quantum system (the
Berezinski-Kosterlitz-Thouless (BKT) transition in
the two-dimensaional (2D) XY model), like in the
usual Bose-Hubbard model. (B) The MI-DW transition should be in
the 2D, $Z_2$ (Ising) universality class. (C) The SF-DW transition
involves two different order parameters, so it should have both
BKT and 2D Ising characters. These expectations have, indeed, been
confirmed in a previous work~\cite{rvpai0544}.

The phase transitions involving the new HI phase, which we study
here, have not been accurately characterized so far.  The HI
phase has no local order parameter, so the transitions into this
phase, from the other phases, may be exotic. We expect the HI-DW
transition to have 2D Ising character, because the DW phase has
long-range order, with a well-defined local order parameter. We
might, at first sight, expect the SF-HI transition to be of the
BKT type; but the interplay of quasi-long-range order, in the SF
phase, and the  string correlations, in the HI phase, might lead
to an exotic transition. The MI-HI transition is perhaps the most
interesting transition in the EBH model, because both phases have
gaps and no local order parameter; in addition, the HI phase is
characterized by a string order parameter. The field theory of
this transition has been studied in Ref.~\cite{berg0844}; and a
similar transition, in a spin-1 model, has been studied
numerically in Ref.~\cite{gaussian144}. This transition has been
argued to be in the Gaussian universality class, which has
continuously varying critical exponents and was first studied in
models of roughening~\cite{ALuther}. In this work, we perform
extensive numerical studies to elucidate the natures of these
transitions.

\begin{figure*}
\centering \includegraphics[width=18cm,height=8cm]{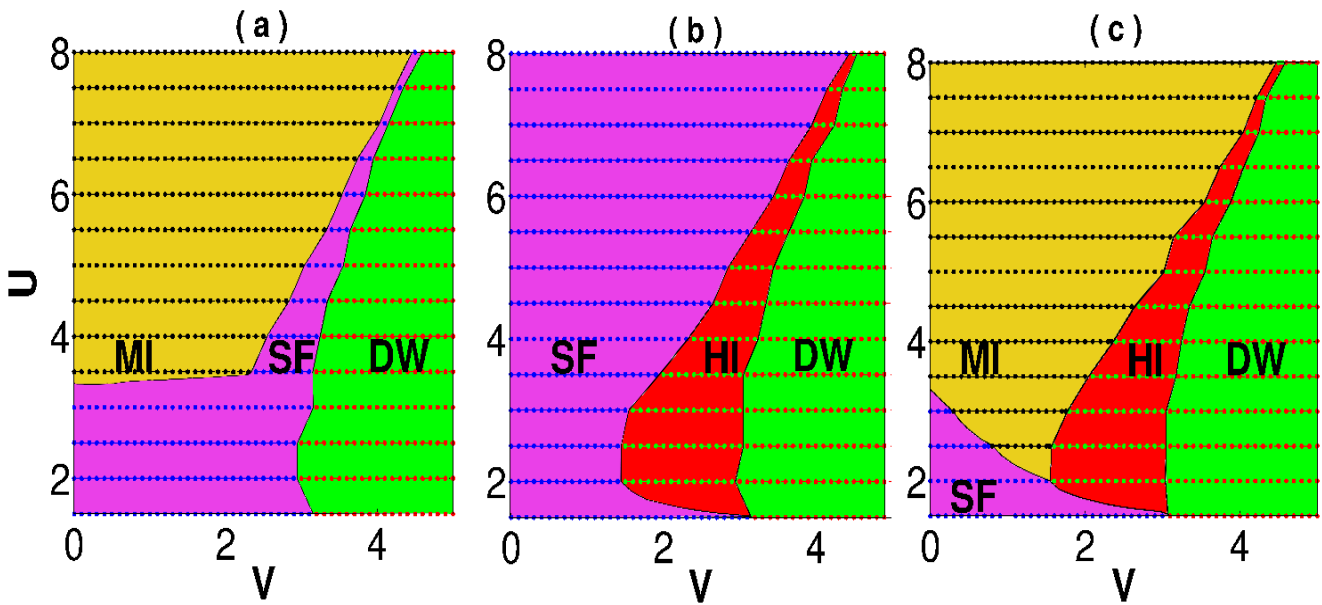}
\caption{\label{fig:PhaseD}(Color online) {Phase diagrams in the $(U,V)$ plane for the
1D EBHM with the following constraints $(a)$ T1 $(N = L)$,
$(b)$ T2 $(N = L + 1)$, and $(c)$ T3 $(N = L)$ and $\mu_r=-\mu_l=2$ showing
Mott-insulator (MI ochre), superfluid (SF purple), Haldane-insulator
(HI red), and density-wave (DW green) phases and the phase boundaries
between them; in this range of $U$ and $V$ all transitions are continuous;
at larger values of $U$ and $V$ the MI-DW and HI-DW transitions become
first-order.}}
\end{figure*}

We present density-matrix-renormalization-group
(DMRG)~\cite{rvpai0544} studies of the phases, transitions, and
the role of boundaries in the 1D EBHM with a mean filling of one
boson per site. Although a few DMRG
studies~\cite{rvpai0544,torre0644,berg0844,deng1244,rossini1244}
have been carried out earlier, none of them has elucidated
completely clearly the universality classes of the continuous
transitions in the 1D EBHM; nor have they compared in detail, the
phase diagrams of this model with different types of boundaries
and filling. Our study, which has been designed to obtain the
universality classes of these transitions and to examine the
roles of boundary states and filling in stabilizing different phases, yields
a variety of interesting results that we summarize qualitatively
below: If the filling $\rho=1$, in the thermodynamic limit, we
obtain three types of phase diagrams, P1
(Fig.~\ref{fig:PhaseD}(a)), P2 (Fig.~\ref{fig:PhaseD}(b)), and P3
(Fig.~\ref{fig:PhaseD}(c)), for the following three different types of
constraints on the system: T1 (number of bosons $N=L$, where $L$
is the number of sites); T2 ($N=L+1$); and T3 ($N=L$ and
additional chemical potentials at the boundary sites, as we
describe below). For P1 we obtain SF, MI, and DW
phases and we confirm, as noted earlier~\cite{rvpai0544}, that
the SF-MI transition is in the
BKT~\cite{kogutreview44} universality class and the SF-DW has
both BKT and 2D Ising characteristics; at
sufficiently large values of the repulsive parameters, the MI-DW
transition is of first order~\cite{rvpai0544}. The phase diagram
P2 displays SF, HI, and DW phases, with continuous SF-HI and
HI-DW transitions in BKT and 2D Ising universality classes,
respectively.  The phase diagram P3 has SF, MI, HI, and DW
phases, with SF-MI, MI-HI, HI-DW in BKT, 2D
Gaussian~\cite{gaussian144}, and 2D Ising universality classes,
respectively; the SF-HI transition is more exotic than a simple
BKT transition; at sufficiently large values of the repulsive
parameters, there is a first-order MI-DW phase boundary.

 The remaining part of this paper is organized as
follows. In Sec.~\ref{sec:model} we give details of the EBH model and define the quantities we calculate. Section~\ref{sec:results} is devoted to our results. Sec.~\ref{sec:Conclusions} contains concluding remarks. The details of the DMRG we use are given in the Appendix.

\section{Model}\label{sec:model}

We set the scale of energies by choosing $t = 1$. We work
with a fixed value of the filling $\rho$ and focus on $\rho=1$,
i.e., a filling of one boson per site in the thermodynamic limit,
which we realize in three different ways, to highlight the role
of filling and boundary states, by using the three different constraint
conditions T1, T2, and T3.  In all these three cases, we have
open boundaries; in T3, we include boundary chemical potentials
$\mu_l=-2t$ and $\mu_r=2t$ at the left and right boundaries,
which are denoted by the subscripts $l$ and $r$, respectively. We study
this model by using the DMRG method that has been described in Ref.\cite{rvpai0544} (see also the Appendix). To characterize the phases and transitions
in this model, we calculate the following thermodynamic and correlation
functions and the entanglement entropy (familiar from quantum information):
\begin{eqnarray}
&&R_{dw}(\mid i -
j\mid)=(-1)^{\mid i-j\mid}\langle \delta {\hat n}_{i}\delta {\hat
n}_{j} \rangle,\nonumber \\
&&R_{string}(\mid i - j \mid)=\langle
\delta{{\hat n}_{i}} e^{({i\pi \sum_{l=i}^{j}\delta {{\hat
n}_{l}}})} \delta {{\hat n}_j} \rangle,\nonumber \\
&&R_{SF}(\mid i
- j \mid)=\langle a_{i}^{\dagger} a_{j}\rangle, \nonumber \\
&&\mathcal{S}_\pi = \sum_{i,j=1}^{L}e^{i\pi(i-j)}\langle {\hat
n}_{i}{\hat n}_{j} \rangle/L^2,\nonumber \\
&&n(k=0) = \sum_{i,j}^{L}\langle a_{i}^{\dagger} a_{j}\rangle/L,\nonumber \\
&&G_{L}^{NG} = E_{L}^{1}(N)-E_{L}^{0}(N),\nonumber \\
&&G_{L}^{CG} = E_{L}^{0}(N+1)+E_{L}^{0}(N-1)-2E_{L}^{0}(N),\nonumber \\
&&\xi_{L} = \sqrt{\frac{\sum_{i,j=1}^{L}(i-j)^2\langle
a_{i}^{\dagger} a_{j}\rangle} {\sum_{i,j=1}^{L}\langle
a_{i}^{\dagger} a_{j}\rangle}} ,\nonumber \\
&&S_L(l) = -\sum_{i}^{n_{states}}\lambda_i \log_2\lambda_i, \nonumber \\
&&F = LG_{L}^{CG}\left(1+\frac{1}{2\ln L+B}\right), \nonumber\\
&&D = \ln L-a|V-V_{c}|^{-1/2},
\label{eq:measure}
\end{eqnarray}
where  $\delta{\hat n}_i={\hat n}_i-\rho$, the correlation
functions for the DW, HI, and SF phases are $R_{dw}(|i-j|)$,
$R_{string}(|i-j|)$, $R_{SF}(|i-j|)$, respectively,
$\mathcal{S}_{\pi}$ is the density-density structure factor at
wave number $k=\pi$, $n(k=0)$ the momentum distribution of the
bosons at $k=0$, $G_{L}^{NG}$ and $G_{L}^{CG}$ are neutral and
charge gaps~\cite{footnote44}, $E_{L}^{0}(N)$ and $E_{L}^{1}(N)$
are the ground-state and first-excited-state energies,
respectively, for our system with $L$ sites and $N$ bosons,
$\xi_{L}$ is a system-size-dependent correlation length; $S_L(l)$
is the von-Neumann block entanglement entropy, $\lambda_i$, the
eigenvalues of the reduced density matrix for the right block, of
length $l$, which yield the entanglement spectrum in our
DMRG~\cite{rvpai0544}, and $n_{states}$ is the number of states
that we retain for this density matrix; note that $S_L(l) =
c\lambda$ at a critical point~\cite{kollath44}, with $c$ the
central charge and the log-conformal distance
$\lambda=(\log[\frac{2L}{\pi}\sin(\frac{\pi l}{L})])/6$; $a$ and
$B$ depend on the critical value $V_{c}$, at which the concerned
transition occurs; $V_{c}$ depends on $U$. The order parameters
for the DW and HI phases are
$O_{dw}=\sqrt{\lim_{|i-j|\to\infty}R_{dw}(\mid i - j\mid)}$ and
$O_{string}=\sqrt{\lim_{|i-j|\to\infty}R_{string}(\mid i -
j\mid)}$, respectively; for the DW, HI and SF phases, we also use
the Fourier transforms of the correlation functions, which we
denote by $R_{dw}(k)$, $R_{string}(k)$, $R_{SF}(k)$,
respectively. Our calculations have been performed with $100 \leq
L \leq 300$, a maximal number $n_{max}=6$ of bosons at a site,
and up to $n_{states}=256$ states in our density matrices; we
have checked, in representative cases, that, for the ranges of
$U$ and $V$ we cover, our results are not affected significantly
if we use $n_{max}=4$ and $n_{states}=128$; for the MI-HI
transition we have also used $L=1024$.

\begin{figure*}[htbp]
\centering \includegraphics[width=18cm,height=8cm]{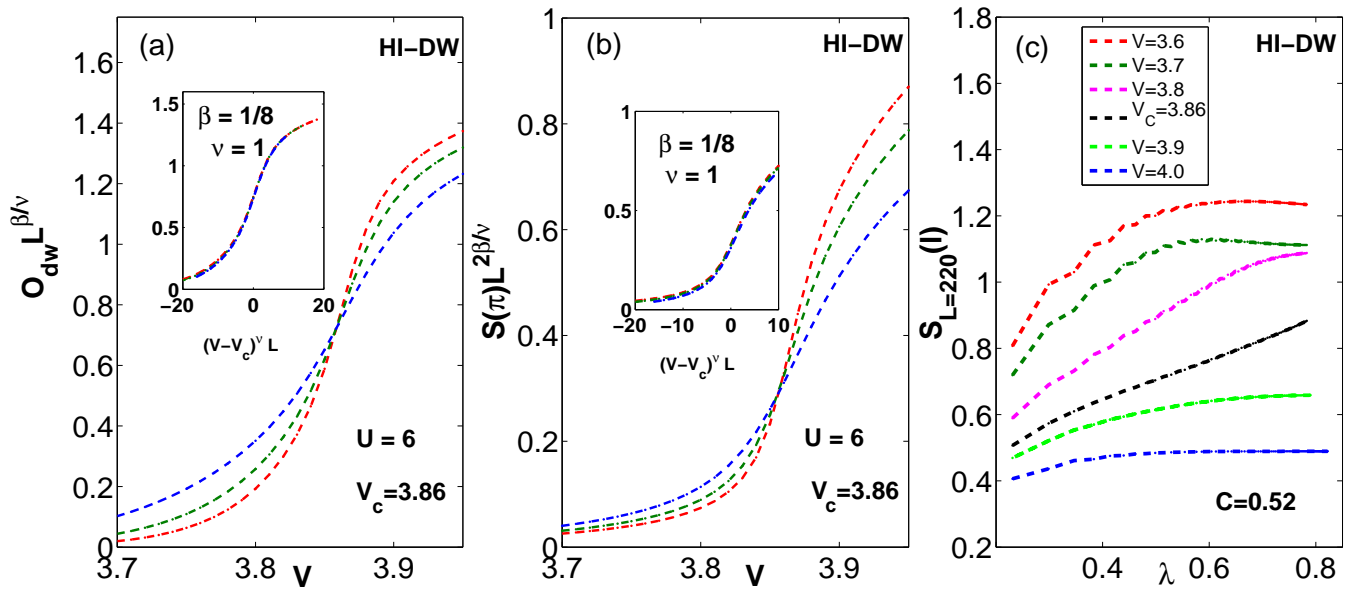}
\caption{\label{fig:OPS} (Color online)
Plots for the HI-DW transition at $U=6$, with $L=200$ (red
curve), $L=150$ (green curve), and $L=100$ (blue curve), of
$(a)$ the scaled DW order parameter $O_{dw}L^{\beta/\nu}$
versus $V$ (the inset shows $O_{dw}L^{\beta/\nu}$ versus
$(V-V_{c})^{\nu}L$), where $\beta$ and $\nu$ are order-parameter
and correlation-length exponents, $(b)$ the $k=\pi$ scaled
structure factor $\mathcal{S}{({\pi})}L^{2\beta/\nu}$ versus $V$
(the inset shows $\mathcal{S}{({\pi})}L^{2\beta/\nu}$ versus
$(V-V_{c})^{\nu}L$), and  $(c)$ the block entanglement entropy $S_L(l)$
versus the logarithmic conformal distance
$\lambda={\frac{1}{6}}\log[\frac{2L}{\pi}sin(\frac{\pi l}{L})]$
for an open system of length $L = 220$ for different values of
$V$ (the slope of the linear part of the curve for $V_c=3.86$ yields
a central charge $c = 0.52$). The system is of type T3.}
\end{figure*}

\section{Results}
\label{sec:results}

In Figs.~\ref{fig:PhaseD} (a), (b), and (c) we depict our DMRG
phase diagrams, in the $(U,V)$ plane, for the 1D EBHM with
constraints T1, T2, and T3, respectively. These phase diagrams
show MI (ochre), SF (purple), HI (red), and DW (green) phases and
the phase boundaries between them; the points inside these phases
indicate the values of $U$ and $V$ for which we have carried out
our DMRG calculations. In the ranges of $U$ and $V$ that we
consider here, all transitions are continuous; at larger values
of $U$ and $V$ than those shown in Fig.~\ref{fig:PhaseD}, the
MI-DW transitions become first-order, as shown for T1 in
Ref.\cite{rvpai0544}; the direct MI-DW transition, at large values
of $U$ and $V$, is also of first order with the T3 constraint.
Note that MI (HI) phases appear with conditions T1 and
T3 (T2 and T3). The MI phase occurs at a strict commensurate
filling (T1 with $N=L$) and is characterized by a charge gap,
which arises because of the onsite interaction $U$ that penalizes
the presence of more than one boson at a site. With the
constraint T2, $N=L+1$, so there is one more boson in the system
than in case T1; this extra boson can hop over the commensurate,
T1, MI state, produce gapless excitations in the upper
Mott-Hubbard band, and thereby destroy the MI phase, which is
replaced by SF or HI phases.

The origin of the HI phase in the 1D EBHM has been elucidated in
the Introduction. This phase also has gapless edge modes, which
destroy the charge gap at strict commensurate filling (T1).
However, with an extra boson (T2), these gapless modes get
quenched, so we obtain an HI phase with a gap. At strict
commensurate filling, we can gap out the edge modes by applying
boundary chemical potentials $\mu_l$ and $\mu_r$ (T3). This
allows us to obtain the HI phase and also the MI phase. A similar
procedure was used in the study of integer spin
chains~\cite{WITHHUSE1993}. Note also that the HI phase exists
because of boson-number fluctuations about the Mott state; so we
expect the HI to give way to the MI phase at large values of $U$
at which such number fluctuations are suppressed, in much the
same way as the SF yields to the MI at large values of $U$. In
the large-$U$ and large-$V$ regions, which we do not show in the
phase diagrams of Fig.~\ref{fig:PhaseD}, there is a direct,
first-order, MI-DW transition~\cite{rvpai0544}.

\begin{figure*}[htbp]
\centering \includegraphics[width=18cm,height=8cm]{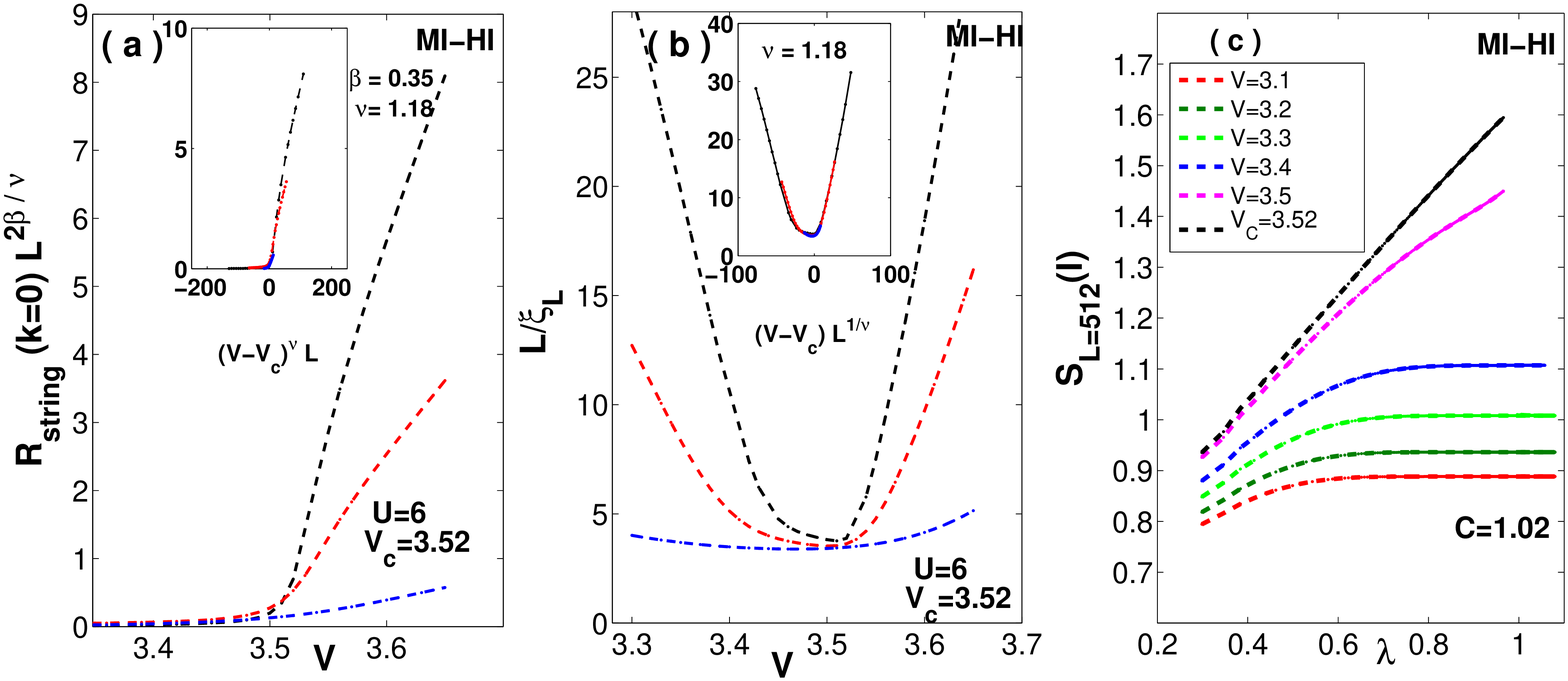}
\caption{\label{fig:OPS2} (Color online)
For this particular phase transition, we used lengths up to $L=1024$
to calculate the critical point, central charge, and critical
exponents; the values for these exponents do not change significantly
if we use lengths up to $L=290$.
$(a)$-$(c)$ show the analogs,
for the MI-HI transition, of the plots in Figs. 2 $(a)$-$(c)$ for the HI-DW transition but
with $(a)$ $O_{dw}L^{\beta/\nu}$ replaced by
$R_{string}(k=0)L^{2\beta/\nu}$, the scaled Fourier transform of
string correlation function at $k=0$, and $(b)$
$\mathcal{S}{({\pi})}L^{2\beta/\nu}$ replaced by $L/\xi_L$;
in (a) and (b) the values of $L$ are $1024$ (black curve) $512$ (red curve),
and $128$ (blue curve). The system is of type T3.}
\end{figure*}

\begin{figure*}[htbp]
\centering \includegraphics[width=18cm,height=8cm]{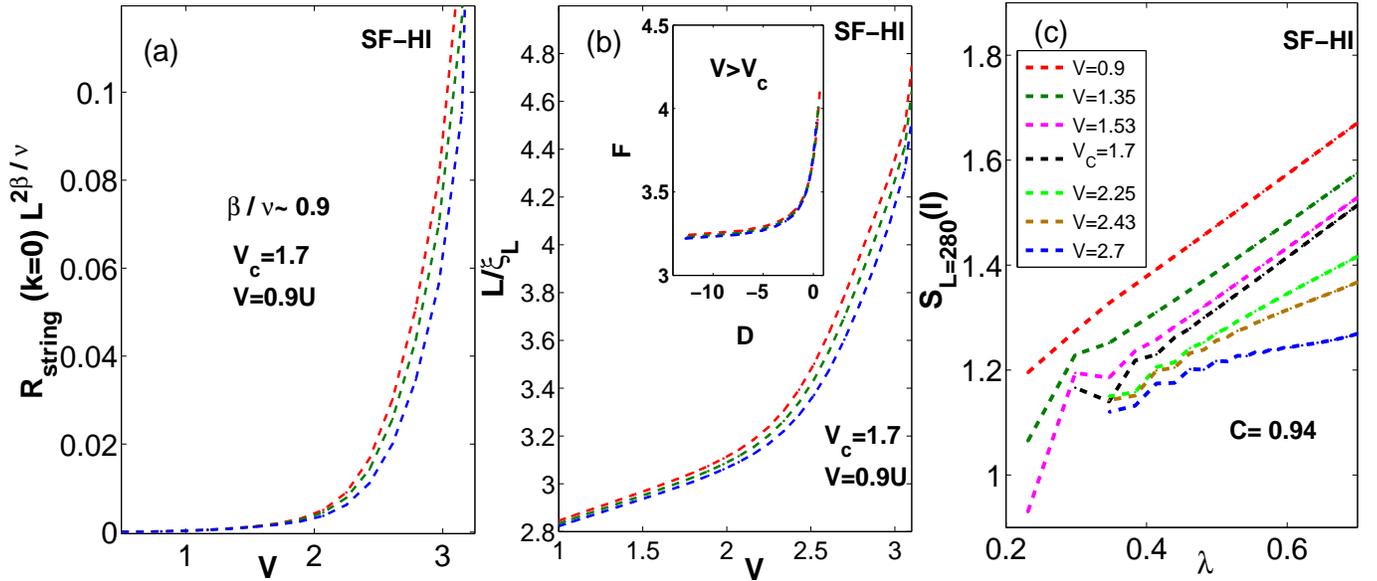}
\caption{\label{fig:OPS3} (Color online)
$(a)$-$(c)$ show the analogs, for the SF-HI transition,  of the plots in Figs. 3 $(a)$-$(c)$ for the MI-HI transition(the inset in $(b)$ shows a plot
of the scaled charge gap $F$ versus $D$ (see
Eq.~\ref{eq:measure}); in (a) and (b) the values of $L$
are $280$ (red curve) $240$ (green curve), and $200$ (blue curve). The system is of type T3.}
\end{figure*}

\begin{figure*}[htbp]
\centering \includegraphics[width=18cm,height=8cm]{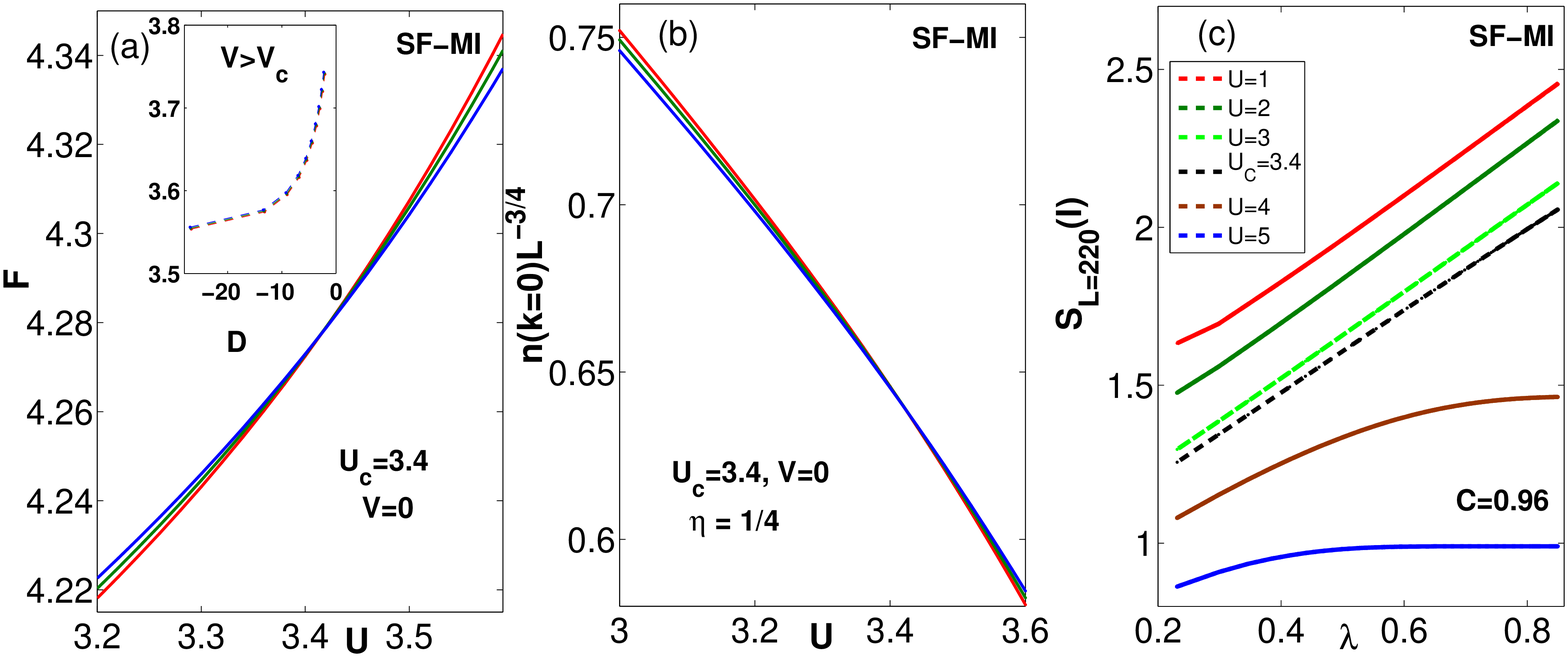}
\caption{\label{fig:OPS4} (Color online)
$(a)$-$(c)$ show the analogs, for
the SF-MI transition,  of the plots in Figs. 3 $(a)$-$(b)$ for the MI-HI transition but with
$(a)$ $R_{string}(k=0)L^{2\beta/\nu}$ replaced by the scaled
charge gap $F$ and $V$ replaced by $U$ and $(b)$ $L/\xi_L$
replaced by $n(k=0)L^{-3/4}$; in (a) $L$ is $200$ (red curve) $180$
(green curve), $160$ (blue curve) and in (b) $L$ is $250$ (red curve)
$230$ (green curve), and $210$ (blue curve). The system is of type T3.}
\end{figure*}

We now obtain the universality classes of the continuous
transitions in the phase diagrams of Fig.~\ref{fig:PhaseD}. It
has been noted in Ref.\cite{rvpai0544} that the MI-SF transition in
Fig.~\ref{fig:PhaseD} (a), with the constraint T1, is in the
Berezinskii-Kosterlitz-Thouless (BKT) class, whereas the SF-DW
has both BKT and 2D Ising characters, in as much as the SF
correlation length shows a BKT divergence but the DW order
parameter decays to zero with a 2D-Ising, order-parameter
exponent; this transition should be in the universality class of
the Wess-Zumino-Witten $SU(2)_1$ theory like the critical point
that lies between the BKT phase and the antiferromagnetic phase
in the spin$-1/2$, XXZ antiferromagnetic
chain~\cite{shankar9044}.  The MI-SF and SF-DW phase boundaries
in Fig.~\ref{fig:PhaseD} (a) merge, most probably at a bicritical
point, beyond which the MI-DW transition is first
order~\cite{rvpai0544}. We concentrate here on the HI-SF, HI-DW, and
HI-MI transitions, all of which can be found in
Fig.~\ref{fig:PhaseD} (c), with the constraint T3, for which we
present results in Figs.~\ref{fig:OPS}, ~\ref{fig:OPS2},
~\ref{fig:OPS3} and ~\ref{fig:OPS4}; we have checked explicitly
that the universality classes of the HI-SF and HI-DW transitions
are the same for both types of constraints T2 and T3.

In Fig.~\ref{fig:OPS} $(a)$ we plot the scaled DW order parameter
$O_{dw}L^{\beta/\nu}$ versus $V$ near the representative point,
$V_c \simeq 3.86$ and $U=6$, on the HI-DW phase boundary for
$L=200$ (red curve), $L=150$ (blue curve), and $L=100$ (green
curve); the inset gives a plot of $O_{dw}L^{\beta/\nu}$ versus
$(V-V_{c})^{\nu}L$; Fig.~\ref{fig:OPS} $(b)$ presents a similar
plot and inset for the scaled structure factor
$\mathcal{S}{({\pi})}L^{2\beta/\nu}$ at wave number $k=\pi$; in
Fig.~\ref{fig:OPS} $(c)$ we plot the von-Neumann block
entanglement entropy $S_L(l)$ versus the logarithmic conformal
distance $\lambda$ for $L = 220$ and a range of values of $V$ in
the vicinity of $V_c(U=6)=3.86$. The values of the exponents
$\beta$ and $\nu$ and the value of the central charge $c$ that we
obtain from Figs.~\ref{fig:OPS} $(a)-(c)$, namely, $\beta = 1/8,
\, \nu = 1$, and $c=0.52$, are consistent with their values in 2D
Ising model. If we restrict the occupancy to one boson per site,
in the limit $U, V \to \infty$ with fixed $U/V$, the DW state can
be represented as $|\dots 20202020 \dots\rangle$ in the
site-occupancy basis; this state is doubly degenerate and clearly
breaks the translational symmetry of the lattice by doubling the
unit cell; in contrast, the HI state has the translational
invariance of the lattice. Given this double degeneracy of the DW
state, we expect that, if the HI-DW transition is continuous, it
should be in the 2D Ising universality class; as we have shown
above, our DMRG results for the HI-DW transition in the 1D EBHM
are consistent with this expectation. Note that the string order
parameter $O_{string}$ is nonzero in both the HI and DW phases;
thus, it is not the order parameter that is required for
identifying the universality class of the HI-DW transition. We
find, furthermore, the string-order-parameter correlation length
is so large in the HI phase that, given the values of $L$ in our
study, it is not possible to get a reliable estimate for this
correlation length.

To investigate the critical behavior of the MI-HI transition, we
plot  $R_{string}(k=0)L^{2\beta/\nu}$ versus $V$, in
Fig.~\ref{fig:OPS2} $(a)$, whose inset shows a plot of
$R_{string}(k=0)L^{2\beta/\nu}$ versus $(V-{V_c})^\nu L$. These
scaling plots indicate that, for $U=6$, the MI-HI critical point
is at $V_c \simeq 3.5$ with exponents $\beta=0.35$ and
$\nu=1.18$; this value of $\beta$ is consistent with the
Gaussian-model~\cite{gaussian144} (superscript $G$) result
$\beta^G=1/\sqrt{8}$. Our value for $\nu$ follows from the plot
of $L/\xi_L$ versus $V$ in  Fig.~\ref{fig:OPS2} $(b)$, whose
inset shows a plot of $L/\xi_l$ versus $(V-{V_c}) L^{1/\nu}$. For
the central charge we obtain $c=1.02$, which is consistent with
$c^G=1$ given our error bars, from the plot of $S_L(l)$ versus
$\lambda$ for $L = 220$ in  Fig.~\ref{fig:OPS2} $(c)$.  As we
have mentioned above, for large $U$ we can restrict the
occupancies in the 1D EBHM to $0,\, 1,$ and $2$ bosons per site
to obtain a spin-1 model. The analog of the MI-HI transition in
this spin-1 model has been shown to be in the Gaussian
universality class~\cite{gaussian144}, which has a fixed exponent
$\beta = 1/\sqrt{8}$ and central charge $c=1$, but an exponent
$\nu$ that changes continuously along the phase boundary.  We
find indeed, from calculations like those presented in
Figs.~\ref{fig:OPS2} $(a)-(c)$, that $\beta$ and $c$ do not
change as we move along our MI-HI phase boundary in
Fig.~\ref{fig:PhaseD} (c), but $\nu$ does; we obtain, in
particular, $\nu=1.18,\, 1.22$, and $1.36$ for $U=6,\, 5$, and
$4$, respectively.  For this particular phase transition, we have
used lengths up to $L=1024$ to calculate the critical point,
central charge, and critical exponents; the values for these
exponents do not change significantly if we use lengths up to
$L=290$.

The critical behavior of the SF-HI transition follows from the
plot of $R_{string}(k=0)L^{2\beta/\nu}$ versus $V$, in
Fig.~\ref{fig:OPS3} $(a)$, and the plot of $L/\xi_L$ versus
$V$, in  Fig.~\ref{fig:OPS3} $(b)$, whose inset shows a plot of
$F$ versus $D$ (see Eq.~\ref{eq:measure}). These scaling plots
indicate that, for $V=0.9U$, the SF-HI critical point is at $V_c
\simeq 1.7$, where the correlation length diverges with an
essential singularity as it does at a BKT
transition~\cite{kogutreview44}. Thus, we cannot define the
exponent $\nu$; however, we can define the exponent ratio
$\beta/\nu$, which governs how rapidly the string order parameter
vanishes as we approach the SF-HI phase boundary from the HI
phase; for this ratio we obtain the estimate $\beta/\nu \simeq
0.9$. We find the central charge $c=0.94$, which is consistent
with $c=1$ given our error bars, from the plot of $S_L(l)$ versus
$\lambda$ for $L = 280$ in Fig.~\ref{fig:OPS3} $(c)$. The field
theory for this SF-HI transition is likely to be non-standard
because the SF phase has algebraic $U(1)$ order, whereas the HI
phase has a non-local, string order parameter. An example of a
similar transition is found in the phase diagram of the
bilinear-biquadratic spin-1 chain, in which the Haldane phase
(the analog of the HI phase here) undergoes a transition to a
critical phase, as we change the ratio of the bilinear and
biquadratic couplings; this transition is known to be described
by an $SU(3)_1$ Wess-Zumino-Witten theory~\cite{itoi44}. The
gapless phase in our model is quite different from that in the
spin-1 chain; and there is no $SU(3)$ symmetry at the SF-HI
transition in the 1D EBHM. Nevertheless, it is likely that the
SF-HI phase transition in our model is of a non-standard type,
which is similar to, but not the same as, the $SU(3)_1$ theory
mentioned above; this point requires more detailed investigations
that lie beyond the scope of our work.  Note, furthermore, that
the SF-HI transition here is not fine-tuned, unlike the one above
in the spin-1 chain with bilinear and biquadratic couplings, in
so far as it occurs along an entire boundary in the phase diagram
(Fig.~\ref{fig:PhaseD} (c)).

\begin{figure*}[htbp]
\centering \includegraphics[width=17cm,height=10cm]{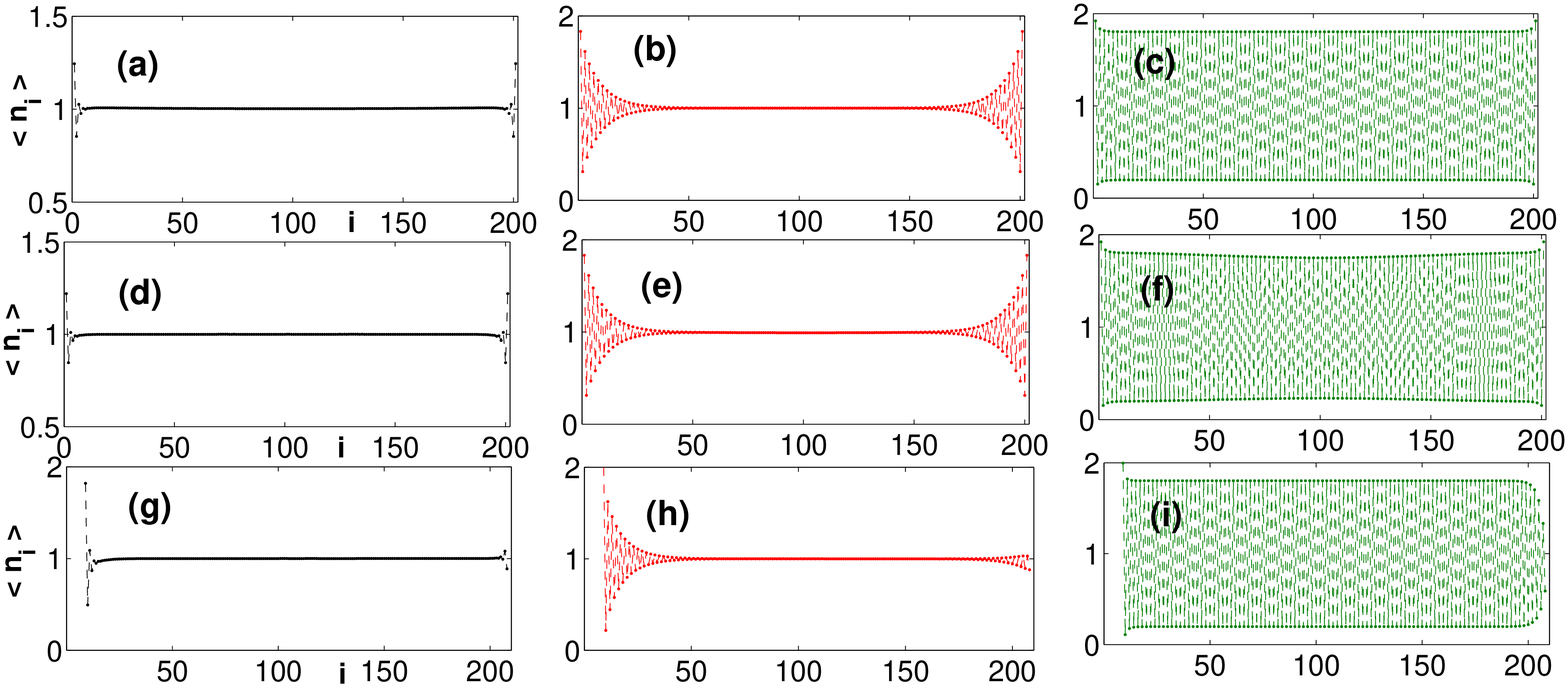}
\caption {(Color online) Plots of $n_i$ versus $i$ for the system
at $U=6$ with the constraints $T_1$ ((a)-(c)), $T_2$ ((d)-(f)),
and $T_3$ ((g)-(i)). In (a), (d) and (g), $V=2.5$, in (b), (e)
and (h) $V=3.7$, and in (c), (f) and (i) $V=4.1$.}
\label{fig:niU6Lp1}
\end{figure*}

\begin{figure*}[htbp]
\centering \includegraphics[width=17cm,height=10cm]{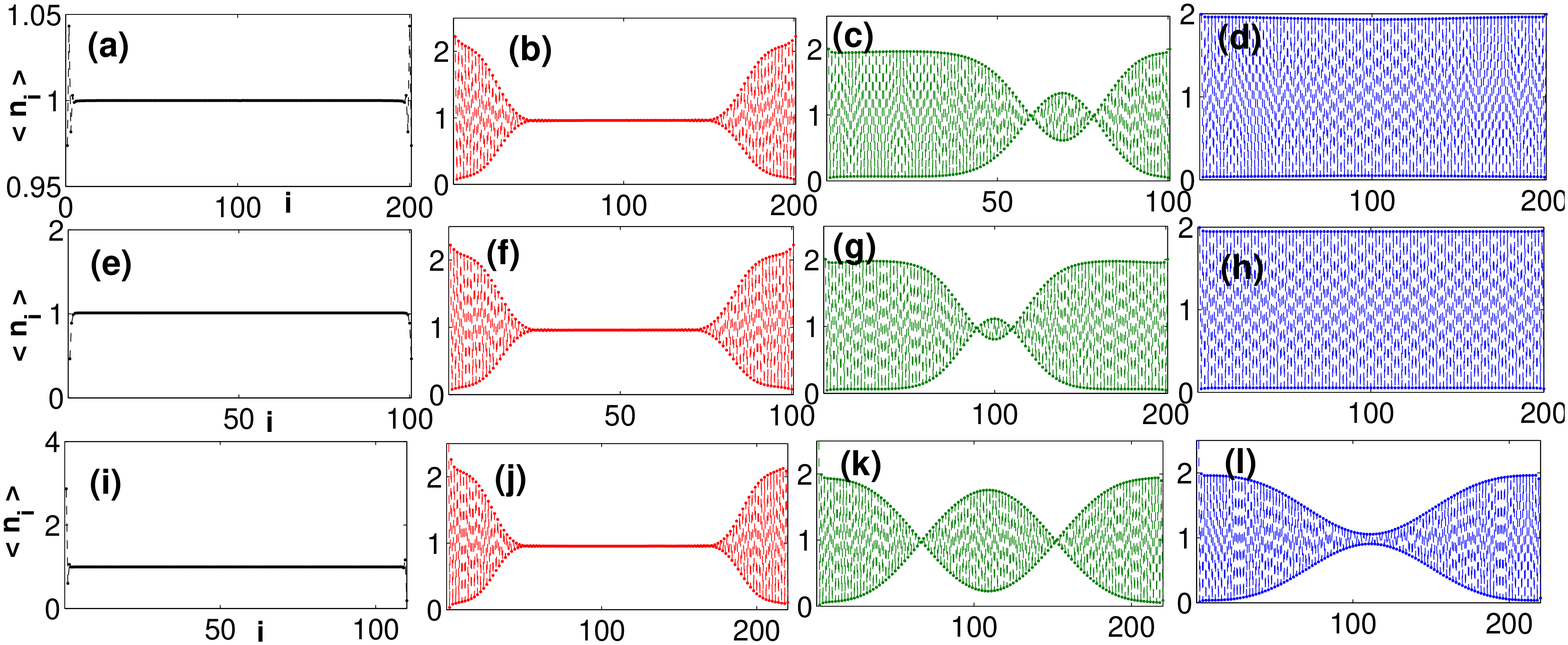}
\caption {(Color online) Plots of $n_i$ versus $i$ for the system
at smaller values of $U (\approx 1)$, with the constraints T1
((a)-(d)), T2 ((e)-(h)) and T3 ((i)-(l)). In (a), (e) and (i)
$V=1$, in (b)-(j), $V=3.5$, in (c)-(k), $V=4.4$ and in (d)-(l)
$V=5$.}
\label{fig:niU1OBCLp1}
\end{figure*}

\begin{figure*}[htbp]
\centering \includegraphics[width=17cm,height=10cm]{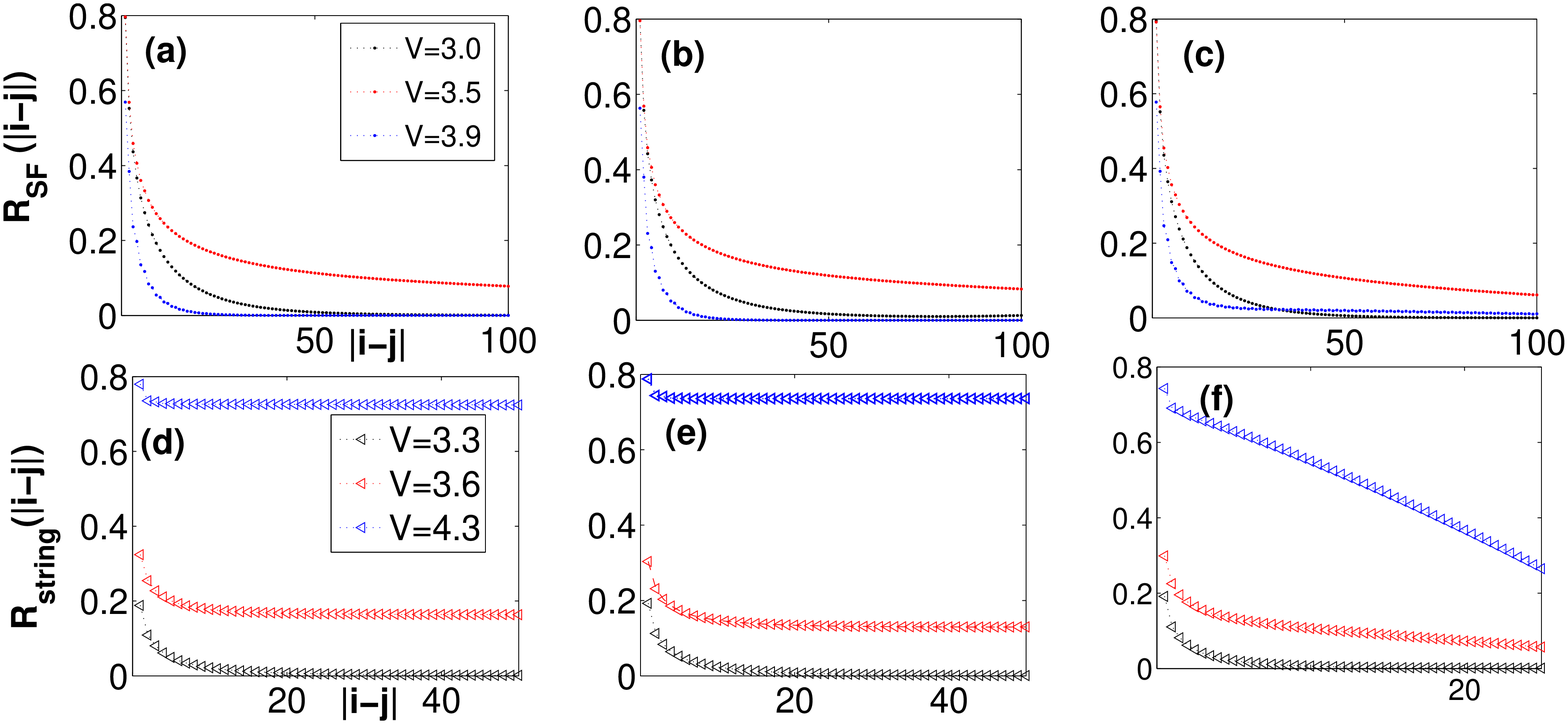}

\caption {(Color online) Illustrative plots of SF ((a)-(c)) and
string correlation functions ((d)-(f)) for the constraints T3
((a) and (d)), T2 ((b) and (e)), and T1 ((c) and (f)), for $U=6$.
}
\label{fig:aiaju6OBCExt}
\end{figure*}

For the sake of completeness, we also show, for the constraint
T3, the BKT nature of the SF-MI transition in
Fig.~\ref{fig:PhaseD} (c), as noted for the T1 case
in~\cite{rvpai0544}, by presenting plots of the scaled charge gap
$F$ versus $D$ (see Eq.~\ref{eq:measure}), the $k=0$ value of
$n(k=0)$, and the block entanglement entropy $S_L(l)$ versus the
log-conformal distance $\lambda$ in Figs.~\ref{fig:OPS4} $(a)$,
$(b)$ and $(c)$, respectively. These plots demonstrate that, at
the MI-SF transition, the correlation length (or inverse charge
gap) has a BKT-type essential singularity (Figs.~\ref{fig:OPS4}
$a$), the SF-correlation-function exponent $\eta=1/4$, and the
central charge is $c=0.96$, which is consistent with $c^{BKT}=1$
given our error bars. It is difficult to give precise error bars
for the exponents and central charges that we have calculated
because there are systematic errors, which include those
associated with the finite values of $n_{max}$ and $n_{states}$;
these errors are hard to estimate.


Our DMRG study can also be used to obtained profiles of the
number of bosons $n_i$ at a site $i$. Illustrative plots are
given in Figs.~(\ref{fig:niU6Lp1}) and (\ref{fig:niU1OBCLp1}) for
a variety of constraints. We expect the boson number to
be constant across the system in the SF, HI and MI phases and to
display a modulation in the DW phase. Fig.~(\ref{fig:niU6Lp1})
shows a plots of $n_i$ vs $i$ for the system at $U=6$ with the
constraints $T_1$, $T_2$ and $T_3$. It can be seen that
$n_i$ is, indeed, constant in the MI phase (away from the
boundaries) and displays modulations in the DW phase; however,
there also appear to be some modulations in SF and HI
phases near the DW phase for the following reasons: (A) proximity to the
DW phase; and (B) the effects of open boundaries (compounded,
in the HI phase, by the presence of edge modes). In
fact, it can be seen that the application of edge potentials to
quench the edge states suppresses the modulation on one side
(with the lower chemical potential) and enhances it on
the other side. The presence of this modulation, even in the SF and HI
phases, prevents a very accurate determination of the phases and
transitions from inspections of plots of $n_i$ versus $i$;
therefore, we require other diagnostics to map out the phase
diagrams accurately.

\begin{figure*}[htbp]
\centering \includegraphics[width=17cm,height=7cm]{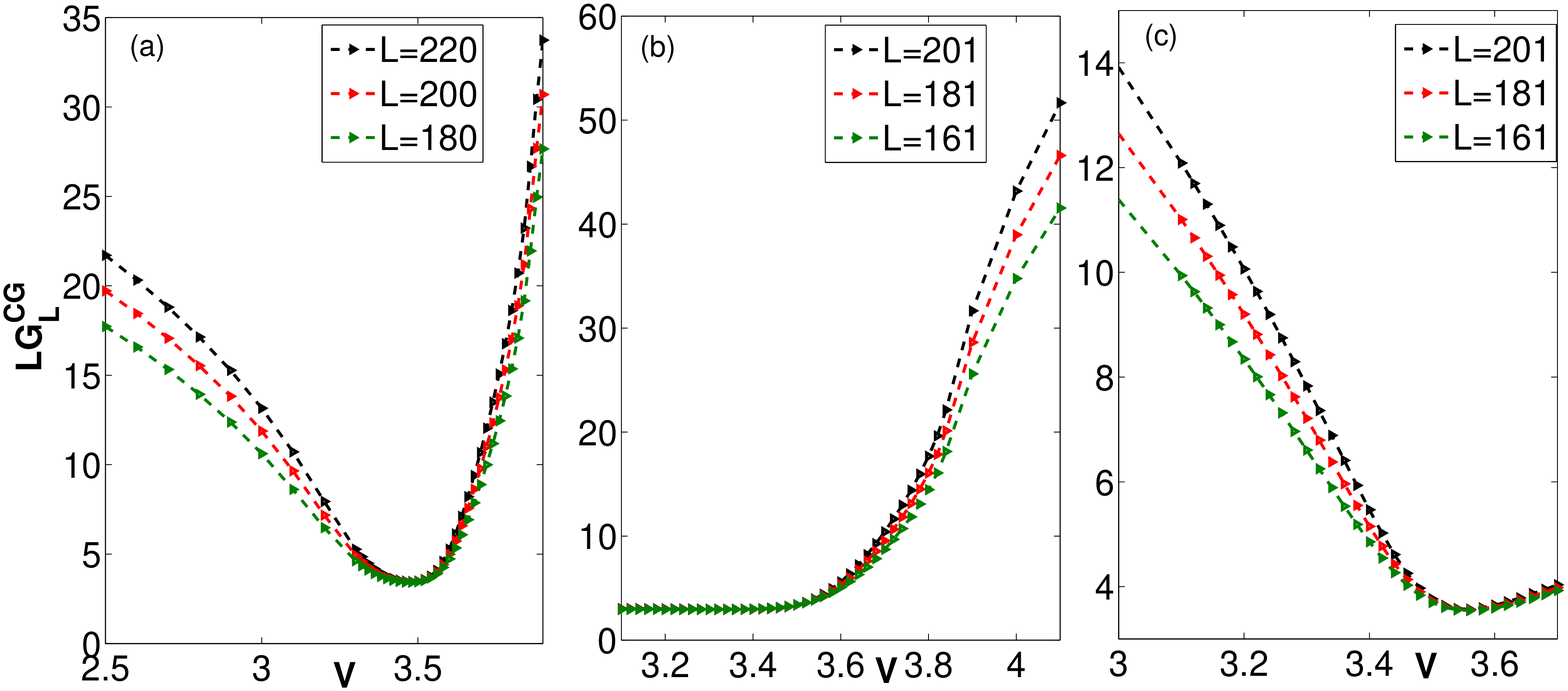}

\caption {(Color online) $L$ times the charge gap for $U=6$ and
different lengths for the system with constraints (a) T3, (b) T2,
and (c) T1. For T3, the gap is non zero in the MI phase at small
values of $V$, vanishes at a point corresponding to the
transition from MI to HI and then opens up again in the HI phase.
For T2, the MI phase does not exist and it is replaced by a gapless
SF at small V, which then gives way to the HI phase, where a gap
opens up. For T1, there is a gapped MI at small $V$, which gives
rise to an SF phase with no gap. The gap closes not just at a
point (for T3) but over a range of values of $V$.
Note that the charge gap does not go to zero at the HI-DW
transition.}
\label{fig:LCGU6OBCEXT}
\end{figure*}

\begin{figure*}[htbp]
\centering \includegraphics[width=17cm,height=7cm]{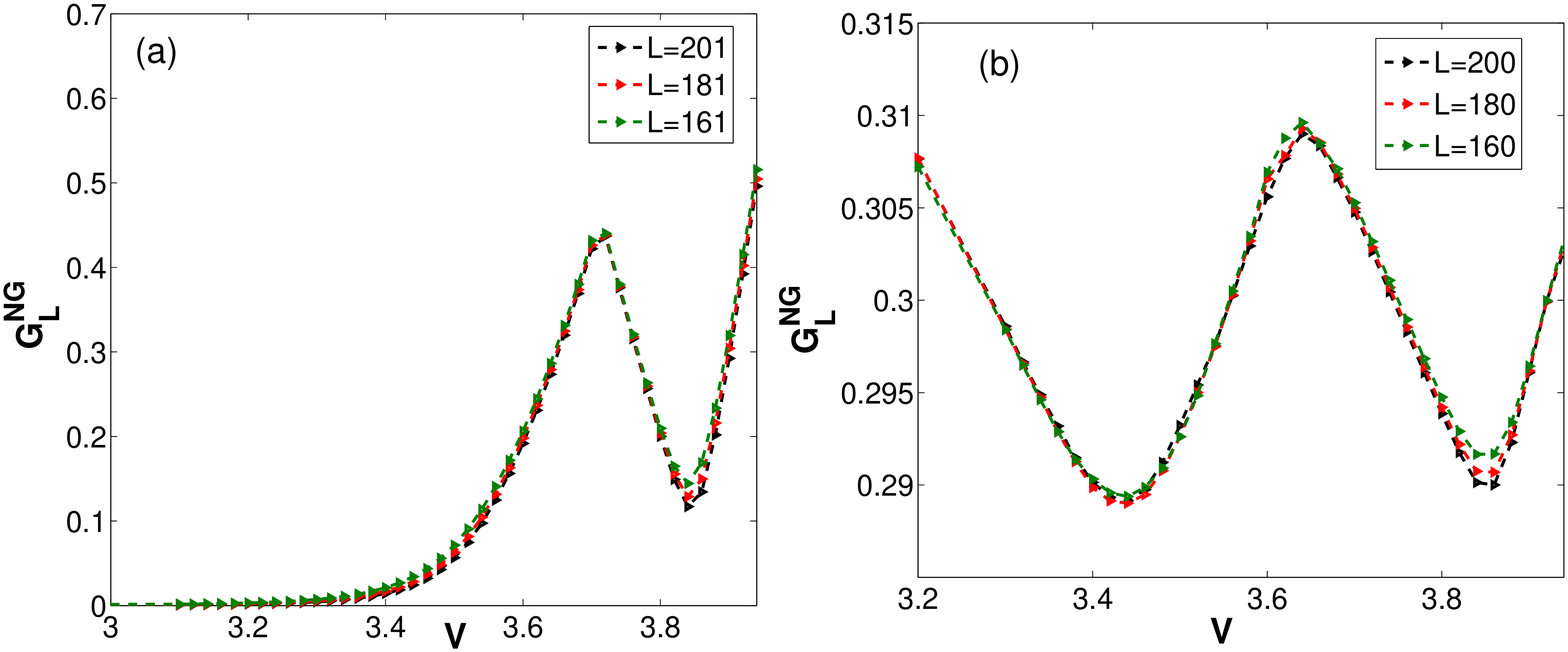}

\caption {(Color online) (a) The neutral gap versus $V$,
for $U=6$, different lengths, and the constraint T2. The
neutral gap is zero in the SF phase and opens up in the HI phase
and, unlike the charge gap, closes at the HI-DW transition and then opens up
again in the DW phase. (b) The neutral gap versus $V$, for
$U=6$, different lengths, and the constraint T3.}
\label{fig:LNGU6OBCEXt}
\end{figure*}

Figure.~(\ref{fig:niU1OBCLp1}) shows that, at small values of $U
(\simeq 1)$, where there are no MI and HI phases, but possibly
an SS phase, the density modulations in the SF phase get stronger
with a region of uniform density sandwiched between regions of
strong modulation. A calculation that considers the entire system
yields, therefore, both nonzero superfluid density and density
modulation. It is tempting to conclude, on the basis of the $n_i$
versus $i$ plots in Fig.~(\ref{fig:niU1OBCLp1}), that the system
displays phase separation between the DW and SF phases and is,
therefore, not in the SS phase.  However, this apparent phase
separation could be an artifact of the open boundaries and system
sizes we employ; hence, we do not make any assertions about the
existence, or lack thereof, of the SS phase. An unambiguous
resolution of this issue requires, perhaps, a DMRG calculation at
fixed values of the chemical potential $\mu$ and not at a fixed
mean density of bosons per site. Such a DMRG calculation lies
beyond the scope of our study here.

We have also obtained various correlation functions in MI, SF,
HI, and DW phases. We give some illustrative plots of SF and
string correlation functions in Fig.(\ref{fig:aiaju6OBCExt}) for
the constraints T1, T2 and T3.  These show that (a) SF
correlations decay exponentially in MI, HI, and DW phases but
algebraically (as power laws) in the SF phase and (b) string
correlations decay to zero exponentially in SF and MI phases, but
asymptote to nonzero values in the HI and DW phases.

In Fig.(\ref{fig:LCGU6OBCEXT}) we show plots of the charge gap
(multiplied by the length $L$ of the system); and in
Fig.(\ref{fig:LNGU6OBCEXt}) we show plots of the neutral gap for
several of the transitions in our model and for different
constraints. At such transitions, one of these gaps vanishes if
the transition is continuous, so the curves for different values
of $L$ (in plots such as those of Figs.(\ref{fig:LCGU6OBCEXT}) -
(\ref{fig:LNGU6OBCEXt})) cross at the transition; this can be
used to detect and characterize these transitions. In particular,
the HI-DW is associated with the vanishing of the neutral gap
alone, whereas HI-MI transitions (Fig.(\ref{fig:LNGU6OBCEXt}))
are associated with the vanishing of the neutral and charge gaps.

%

\begin{figure*}[htbp]
\centering \includegraphics[width=17cm,height=9cm]{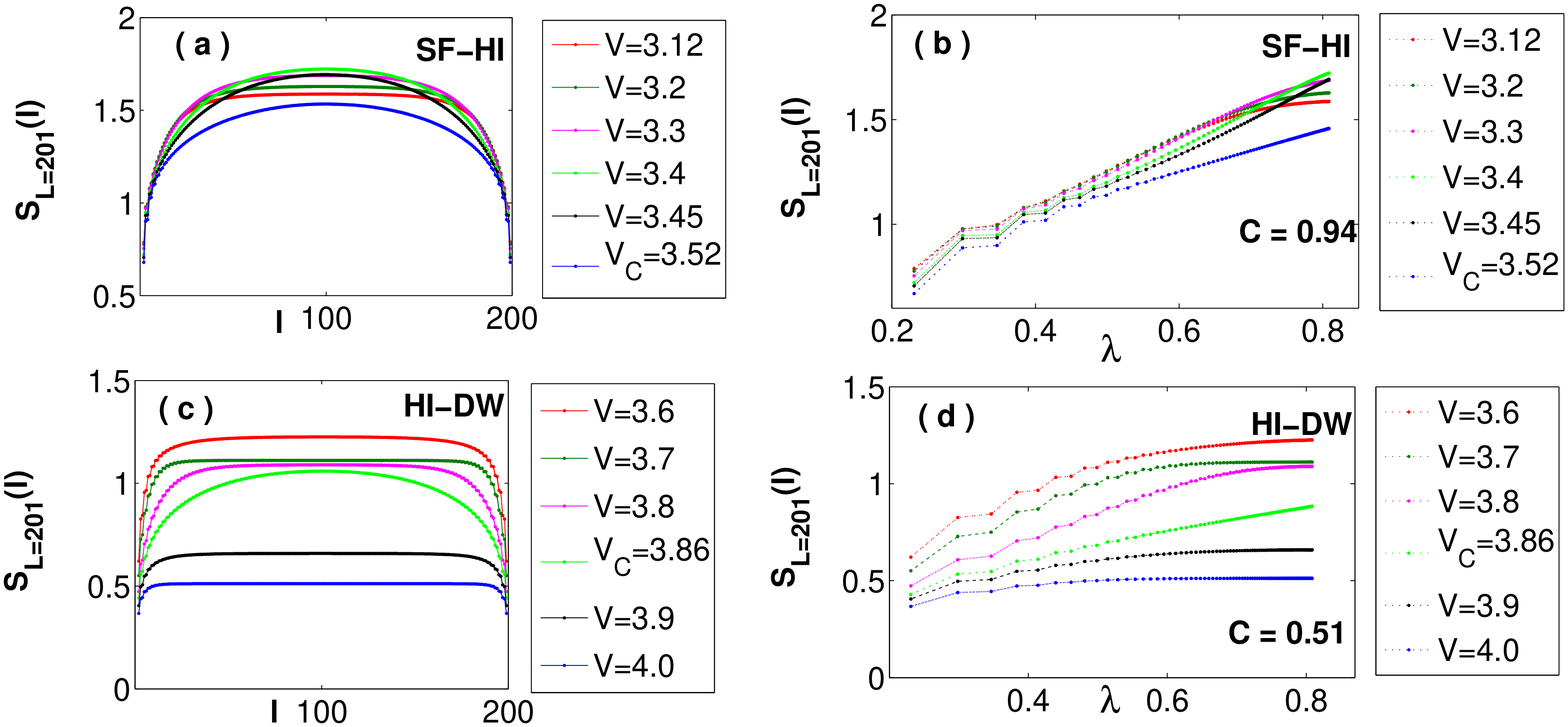}
\caption {(Color online)
Left panels (a) and (c): the static von Neumann block entropy
$S_{L}(l)$ for an open system of length $L = 201$ and a range of
different interaction values $V$, located in the SF-HI and the
HI-DW phases. Right panels (b) and (d): the same block entropies
as a function of the logarithmic conformal distance
$\frac{1}{6}log[(2L/\pi)sin(\pi l/L)]$; the linear behaviors of
the curves for $V = 3.52$ and $3.86$ reveal critical points (see
text) with central charges $c \sim 1$ and $c \sim 0.5$. The rapid
saturation of the entropy for other values of $V$ is a
consequence of the short correlation length in the DW and HI
insulating phases. Here the average filling is 1 and we use the
constraint T2.}
\label{fig:SU6OBCLp1}
\end{figure*}

\begin{figure*}[htbp]
\centering \includegraphics[width=17cm,height=9cm]{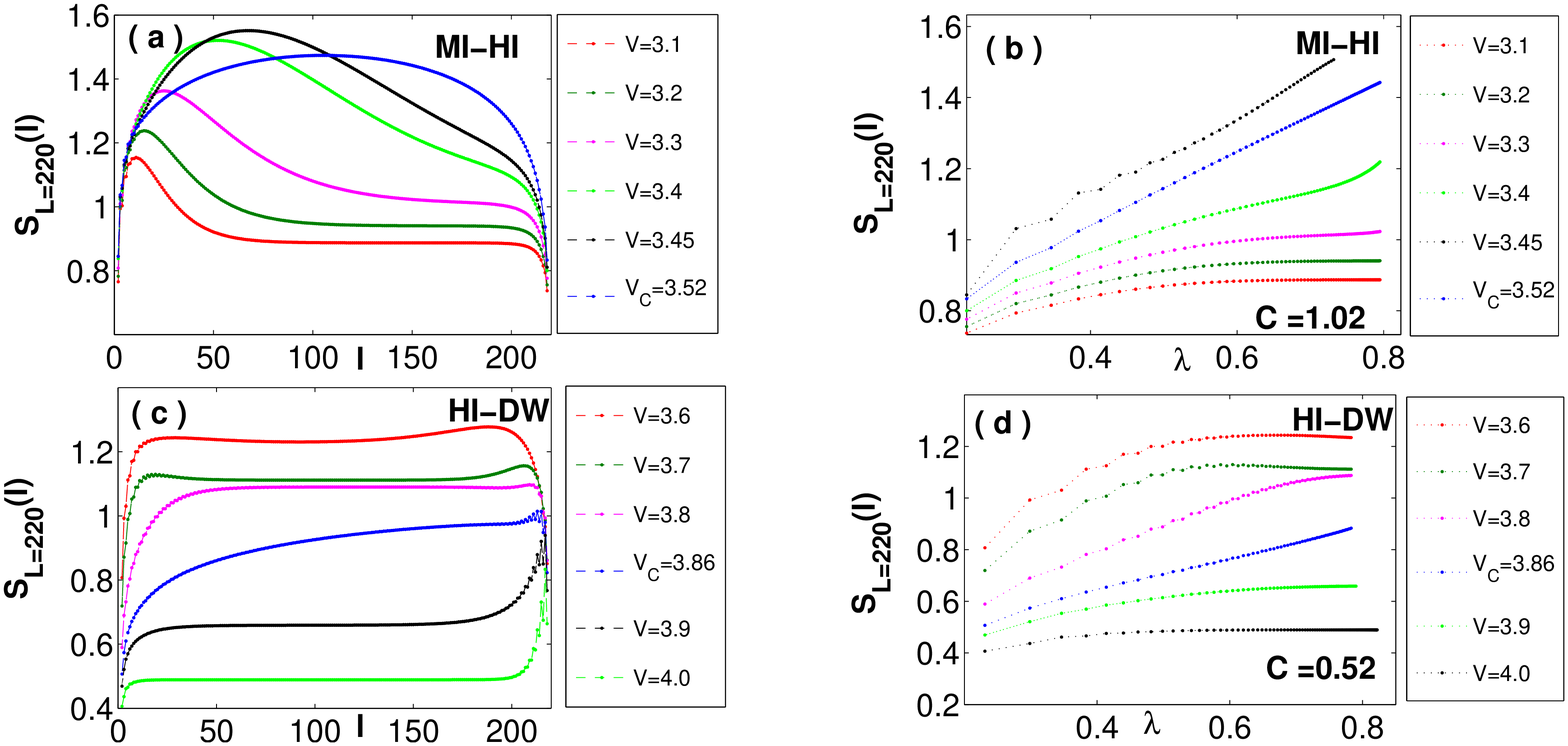}
\caption {(Color online)
Left panels (a) and (c): the static von Neumann block entropy
$S_{L}(l)$ for an open system of length $L = 220$ and a range of
different interaction values $V$, located in the MI-HI and the
HI-DW phases. Right panels (b) and (d): the same block entropies
as a function of the logarithmic conformal distance
$\frac{1}{6}log[(2L/\pi)sin(\pi l/L)]$; the linear behaviors of
the curves for $V = 3.52$ and $3.86$ reveals critical points with
$c \sim 1$ and $c \sim 0.5$. The rapid saturation of the entropy
for other values of $V$ is a consequence of the short correlation
length in the MI and HI insulating phases. Here the average
filling is 1 and we use the constraint T3.}
\label{fig:SU6OBCExterp}
\end{figure*}

\begin{figure*}[htbp]
\centering \includegraphics[width=17cm,height=9cm]{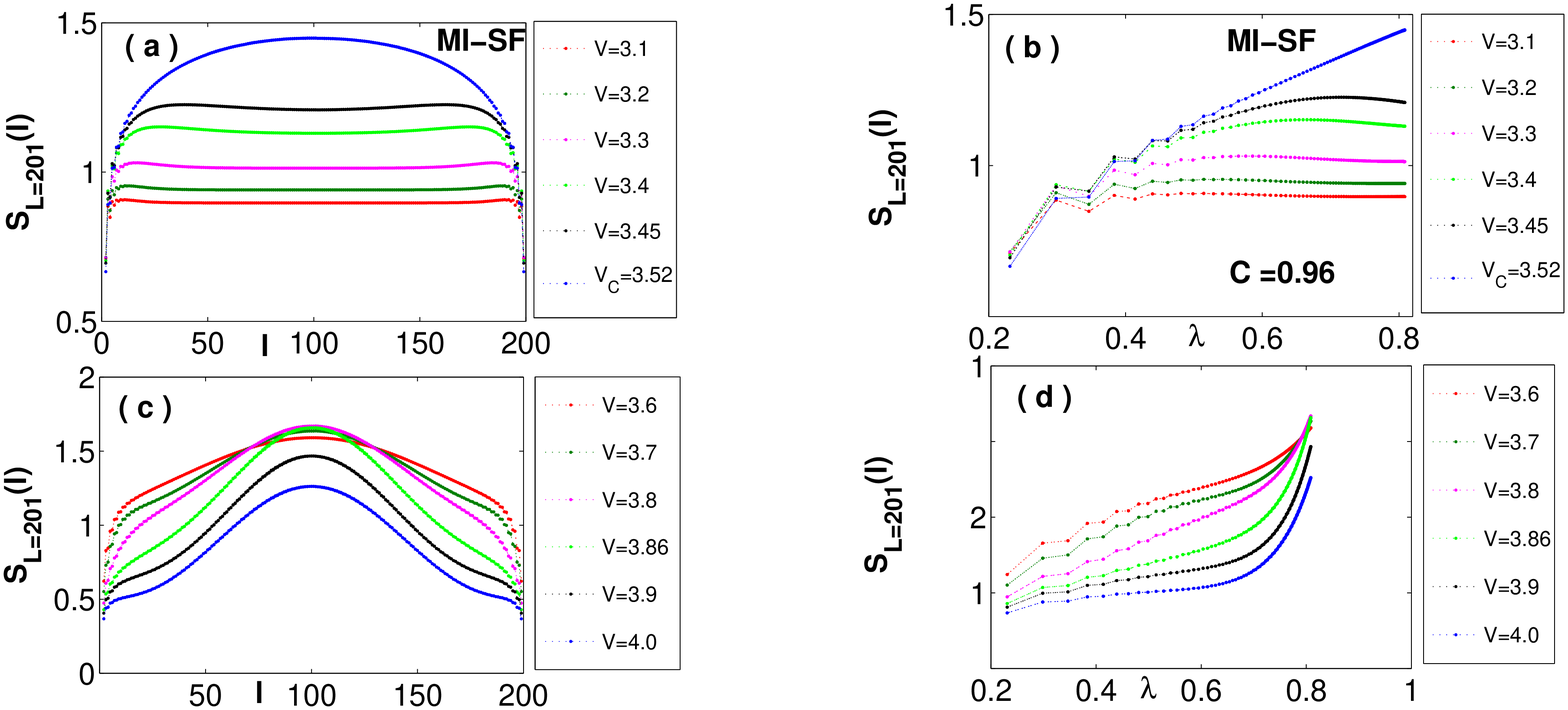}
\caption {(Color online)
Left panel (a): the static von Neumann block entropy $S_{L}(l)$
for an open system of length $L = 201$ and a range of different
interaction values $V$, located in the MI and the SF phases.
Right panels (b): the same block entropies as a function of the
logarithmic conformal distance $\frac{1}{6}log[(2L/\pi)sin(\pi
l/L)]$; the linear behaviour of the curve for $V = 3.52$ reveals
a critical point with central charge $c \sim 1$. The rapid
saturation of the entropy for other values of $V$ is a
consequence of the short correlation length in the MI insulating
phase. Here the average filling is 1 and we use the constraint
T1; (c) and (d) show analogs of the plots in (a) and (b),
respectively, in a parameter range with no continuous
transitions.}
\label{fig:SU6OBCNeL}
\end{figure*}

\begin{figure*}[htbp]
\centering \includegraphics[width=17cm,height=7cm]{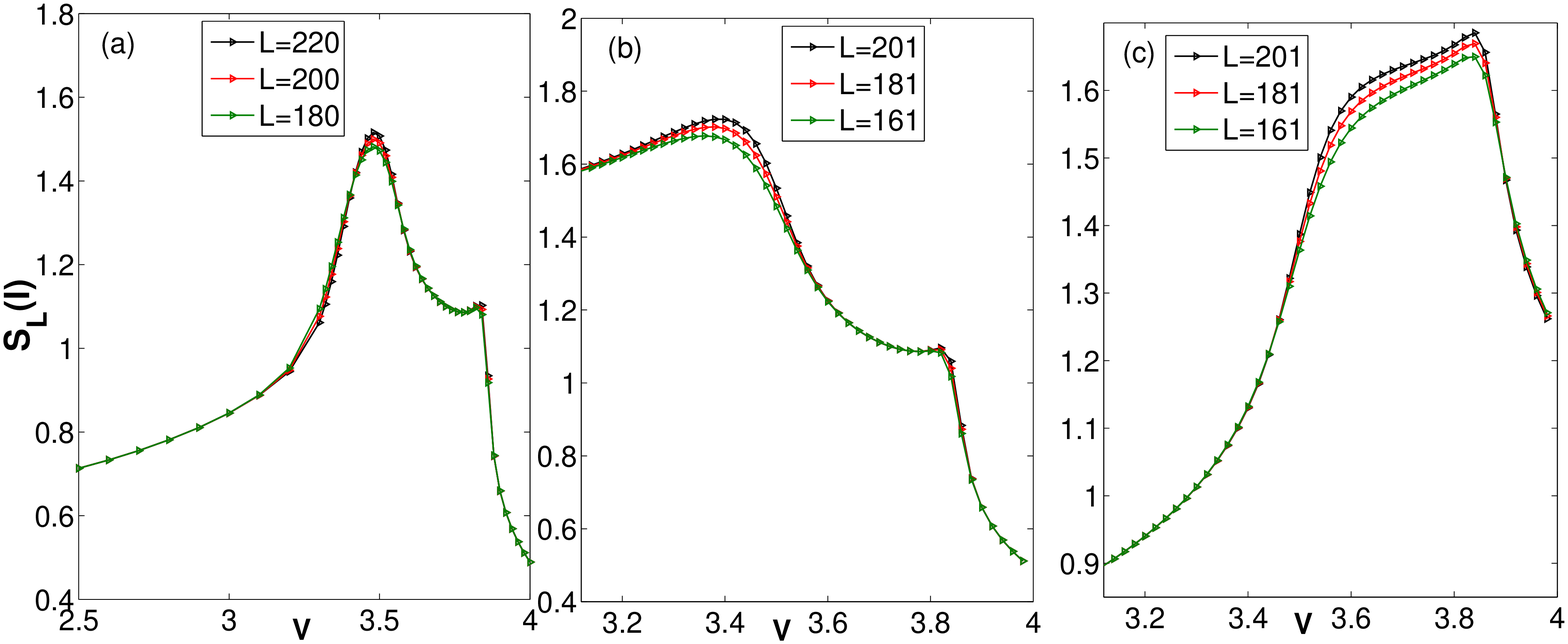}

\caption {(Color online)
In (a) we show the entanglement entropy as a function of $V$ ($l=L/2$); as
$L$ $\longrightarrow \infty$, this shows a jump close to the
Gaussian critical point $V \sim 3.52$, and, in this limit,
the curves for different values of $L$, i.e., 220 (black line),
200 (red line), and 180 (green line), converge to a step function
at the 2D-Ising-type critical point at $V \simeq 3.84$; we use the
constraint T3. (b) and (c) show the analogs, for the constraints T2 and T1,
respectively, of the plot (a) for different lengths $L$ 201
(black line), 181 (red line), and 161 (green line); note
the convergence to a step function, as $L$ increases, at the
2D-Ising-type critical point $V \simeq 3.84$.}
\label{fig:EntropyOBCExU6}
\end{figure*}

We now show representative plots of the static von-Neumann block
entropy $S_L(l)$ (Figs.(\ref{fig:SU6OBCLp1}) -
(\ref{fig:SU6OBCNeL})) versus the the left-block length $l$ (left
panels) and the logarithmic conformal distance $\lambda =
\frac{1}{6}log[(2L/\pi)sin(\pi l/L)]$ (right panels) for a
variety of parameter values and constraints. In the plots versus
the logarithmic conformal distance $\lambda$, linear behaviors of
the curves at certain values of parameters, such as $V$, reveal
critical points; the slopes of these lines yield the central
charge $c$ at the critical point in question. The rapid
saturation of the block entropy versus $\lambda$,
for other values of $V$, is a consequence of the short
correlation length in phases such as the DW and MI phases; if the
correlation length is finite, but long, this saturation can be
slow (as, e.g., in some of the plots in the HI regime). We also
show, for T1-T3 constraints,  plots of the entropy as a
function of $V$; as $L$ $\longrightarrow \infty$, such a curve can
show a jump close to the Gaussian critical point $V \simeq 3.52$;
and, in this limit, the curves for different values of $L$ can
converge to a step function at the 2D-Ising-type critical point
at $V \simeq 3.84$ (see Fig.(\ref{fig:EntropyOBCExU6})).

\section{Conclusions}
\label{sec:Conclusions}
We have performed the most extensive DMRG study of the 1D EBHM
attempted so far. Our work, which extends earlier
studies~\cite{rvpai0544,entspect44,torre0644,berg0844}
significantly, has been designed to investigate the effects of
filling and boundary states on the phase diagram of this model,
to elucidate the natures of the phase transitions here, and to
identify the universality classes of the continuous transitions.
Our study shows that the HI-DW, MI-HI, and SF-MI transitions are,
respectively, in the 2D Ising, Gaussian, and BKT universality
classes; however, the SF-HI transition seems to be more exotic
than a BKT transition in so far as the string order parameter
also vanishes where the SF phase begins. We find that the
different constraints T1, T2, and T3, which lead to the same
value of $\rho$ in the thermodynamic limit, yield different phase
diagrams; this dependence deserves more
attention than it has received so far, especially from the point
of view of the existence of thermodynamic
limits~\cite{thermodynamiclimit44}; a few six-vertex models,
which are known in the statistical mechanics of two-dimensional
models, can show such a boundary-condition dependence in their
bulk phase diagrams as has been shown analytically
in~\cite{Korepin44}. Constraints analogous to the ones that we use have been employed in studies of integer spin chains in Ref.\cite{WITHHUSE1993}. We are not aware of any study of 1D EBHM that compares the constraints T1, T2, and T3. The studies of Ref.\cite{rvpai0544} use the constraint T1 and, therefore, find the phases SF, DW and MI. Ref.\cite{rossini1244} employs constraint T3 and, therefore, obtain the phases SF, DW, MI and HI and Ref.\cite{entspect44} for the same constraint obtain the phases SF, DW, MI, HI and SS. A short version of our study is available at~\cite{tobepublished}.

We hope our work will stimulate more
theoretical and experimental studies of the 1D EBHM and
experimental realizations thereof. 

\section{Acknowledgement}
We thank DST, UGC, CSIR (India) for support and E. Berg, E. G. Dalla Torre, F.
Pollmann, E. Altman, A. Turner, V. Korepin and T. Mishra for useful discussions. RVP and JMK thank, respectively, the
Indian Institute of Science, Bangalore and the Goa University, Goa for hospitality.

\section{Appendix: FSDMRG Calculations}

The Finite-size DMRG (FSDMRG) method has proven to be very useful in
studies of one-dimensional quantum systems~\cite{RVpai244, White44, RVpai344}.
We summarize below the salient featurs of this method. Open boundary
conditions are preferred in this method because the loss of accuracy,
with increasing system size is much less, with these boundary conditions
than in the case of periodic boundary conditions.
The conventional FSDMRG method consists of the following two steps:
\begin{enumerate}
\item The infinite-system, density-matrix renormalization group (DMRG)
method, with which we start, with a system of four sites, add two sites
at each step of the iteration, and then continue until we obtain a
system with the desired number $L$ of sites.
\item The finite-system method in which the system size $L$
is held fixed, but the energy of a target state is improved
iteratively by a sweeping procedure, which we describe below, until
convergence is obtained.
\end{enumerate}
We now follow closely the discussion in Ref.\cite{rvpai0544}.

For a model like the one we consider, we first construct the
Hamiltonian matrix of the superblock configuration
$B^{l}_{1}\bullet\bullet B^{r}_{1}$, where $B^{l}_{1}$ and $B^{r}_{1}$
represent left- and right-block Hamiltonians, respectively; each one of
the $\bullet$ represents a single-site Hamiltonian. In the first step of
the DMRG iteration, both $B^{l}_{1}$ and $B^{r}_{1}$ also represent
single sites, so, at this step, we have a four-site chain. We now
diagonalize the Hamiltonian matrix of the superblock and obtain
the energy and the eigenfunction of a target state. In our study the
\textit{target~state} is the ground state of the system of size $L$,
with either $N = L$ or $N = L \pm 1$ bosons; the latter is required, e.g.,
for obtaining the gap in the energy spectrum. Next, we divide the
superblock into two equal halves, the left and the right parts, which are
treated, respectively, as the \textit{system} and the \textit{environment}.
The density matrix for this \textit{system}, namely,
$B^{l}_{2}\equiv B^{l}_{1}\bullet$, is calculated from the
\textit{target~state}. If we write the \textit{target~state} as
$|\psi\rangle=\sum_{i,j}\psi_{i,j}|i\rangle|j\rangle$, where
$|i\rangle$ and $|j\rangle$ are, respectively, the basis
states of the \textit{system} and the \textit{environment}, then the
density matrix for the \textit{system} has elements
${\rho_{i,i'}}={\sum_{j}}{\psi_{i,j}}{\psi_{{i',j}}}$. The eigenvalues
of this density matrix measure the weight of each of its
eigenstates in the target state. The optimal states for describing
the system are the ones with the largest eigenvalues of the
associated density matrix. In this first step of the DMRG
the superblock, and hence the dimension of the density matrix,
is small, so all the states can be retained. In subsequent steps,
however, when the sizes of the superblocks and density matrices
increase, only the most significant states are retained, e.g., the
ones corresponding to the largest $M$ eigenvalues of the
density matrix (we choose $M=128$ or $256$). We then obtain the
effective Hamiltonian for the \textit{system} $B^l_{2}$ in
the basis of the significant eigenstates of the density matrix;
this is used, in turn, as the left block for the
next DMRG iteration. In the same manner we obtain the
effective Hamiltonian for the right block, i.e.,
$B^{r}_{2}\equiv \bullet B^{r}_{1}$. In the next step of
the DMRG we construct the Hamiltonian matrix for the
superblock $B^{l}_{2}\bullet\bullet B^{r}_{2}$, so the size of
the system increases from $L=4$ to $L=6$. For a system of size $L$,
we continue, as in the first step, by diagonalizing the
Hamiltonian matrix for the configuration
$B^{l}_{(L/2)-1}\bullet\bullet B^{r}_{(L/2)-1}$ and setting
$B^{l}_{L/2}\equiv B^{l}_{(L/2)-1}\bullet$ and
$B^{r}_{L/2}\equiv\bullet B^{r}_{(L/2)-1}$ in the next step
of the DMRG iteration. Thus, at each DMRG step, the left and right
blocks increase in length by one site and the total length $L$ of
the chain increases by 2.

In the infinite-system DMRG method sketched above, the
left- and right-block bases are not optimized in the following
sense: The DMRG estimate for the target-state energy, at the
step when the length of the system is $L$, is not as close to the
exact value of the target-state energy for this system size as it
can be. It has been found that the FSDMRG method overcomes
this problem~\cite{White44}. In this method we first use the
infinite-system DMRG iterations to build up the system to a
certain desired size $L$. The $L$-site superblock configuration
is now given by $B^{l}_{(L/2)-1}\bullet\bullet B^{r}_{(L/2)-1}$.
In the next step of the FSDMRG method, the superblock configuration
$B^{l}_{(L/2)}\bullet\bullet B^{r}_{(L/2)-2}$,  which clearly keeps
the system size fixed at $L$, is used. This step is called
\textit{sweeping} in the right direction since it increases
(decreases) the size of the left (right) block by one site.
For this superblock, the \textit{system} is $B^{l}_{L/2}\bullet$,
the \textit{environment} is $\bullet B^{r}_{(L/2)-2}$, the
associated density matrix can be found, and from its most
significant states the new effective Hamiltonian for the left
block, with $[(L/2)+1]$ sites, is obtained. We sweep again, in
this way, to obtain a left block with $[(L/2)+2]$ sites and so
on until the left block has $(L-3)$ sites and the right block has
1 site, so that, along with the two sites between these blocks,
the system still has size $L$; or, if a preassigned convergence
criterion for the target-state energy is satisfied, this sweeping
can be terminated earlier. Note that, in these sweeping steps,
for the right block we need $B^{r}_{l}$ to $B^{r}_{L-3}$, which
we have already obtained in earlier steps of the infinite-system DMRG.
Next we sweep leftward; the size of the left (right) block
decreases (increases) by one site at each step. Furthermore,
in each of the right- and left-sweeping steps, the energy
of the target state decreases systematically until it converges.

We use a slightly modified form of the FSDMRG method
in which we sweep, as described above, at every step of the
DMRG scheme and not only in the one that corresponds to
the largest value of $L$. Since the superfluid phase in
models such as our model, in $d=1$ and at $T=0$, is critical
and has a correlation length that diverges with the system
size $L$, finite-size effects must be removed by using
finite-size scaling. For this purpose, the energies and
correlation functions, obtained from a DMRG calculation, should
have converged properly for each system size $L$. It is important,
therefore, that we use the FSDMRG method as opposed to the
infinite-system DMRG method, especially in the vicinities of
continuous phase transitions. We find that convergence, to
a specified accuracy for the target-state energy, is faster
in the MI phase than in the SF phase.

The bases of left- and right-block Hamiltonians are
truncated by neglecting the eigenstates of the density matrix
corresponding to small eigenvalues; this leads to truncation
errors. If we retain $M$ states, the density-matrix weight of
the discarded states is $P_{M}=\sum^{M}_{\alpha=1}(1-\omega_{\alpha})$,
where $\omega_{\alpha}$ are the eigenvalues of density matrix.
$P_{M}$ provides a convenient measure of the truncation errors.
We find that these errors depend on the order-parameter correlation
length in a phase. For a fixed $M$, we find very small truncation
errors in the MI and DW phases; these grow as the MI-SF and DW-SF
transitions are approached, and the truncation errors are largest in the
SF phase. In our calculations we choose $M$ such that the truncation
error is always less than $5\times 10^{-6}$; we find that the
values of $M$, mentioned in the main part of this paper, suffice.

The number of possible states per site in the Bose-
Hubbard model is infinite, because there can be any number of
bosons on a site. In a practical DMRG calculation, we must
restrict the number $n_{max}$ of states or bosons allowed per site.
The smaller the interaction parameters $U$ and $V$, the larger
must $n_{max}$ be. As in earlier calculations~\cite{RVpai244,RVpai344}
on related models, we find that $n_{max} = 4$ is sufficient for the
values of $U$ and $V$ considered here; we have checked in
representative cases that our results do not change significantly
if $n_{max} = 5$.

In summary, then, our FSDMRG procedure gives us the energy
$E_{L}(N)$ for the ground state of this model and the associated
eigenstate $|\psi_{0LN}\rangle$. Given these, we can calculate the
energy gaps, order parameters, and correlation functions that
characterize all the phases of this model and thence the phase
diagram.\\



\begin{thebibliography}{70}
\bibitem{footnote44} We use the conventional terminology charge gap
even though we do not have charged bosons in mind. Indeed, some
authors refer to our DW phase as a CDW phase.
\end{thebibliography}
\end{document}